\documentclass[12pt,manuscript]{elsart}
\usepackage{epsfig,harvard,amssymb,rotating,lscape}

\makeatletter
\def\elsartstyle{%
        \def\normalsize{\@setfontsize\normalsize\@xiipt{14.5}}
        \def\small{\@setfontsize\small\@xipt{13.6}}
        \let\footnotesize=\small
        \def\large{\@setfontsize\large\@xivpt{18}}
        \def\Large{\@setfontsize\Large\@xviipt{22}}
        \skip\@mpfootins = 18\p@ \@plus 2\p@
        \normalsize
}
\makeatother

\def\apj{ApJ}
\def\apjs{ApJS}
\def\aap{A\&A}
\def\mnras{MNRAS}

\newcommand{\Alfven}{$\rm Alfv\acute{e}n$}

\newcommand{\D}{\partial}
\newcommand{\DD}{\frac}

\newcommand{\mm }{\mathrm}

\newcommand{\beq}{\begin{equation}}
\newcommand{\eeq}{\end{equation}}
\newcommand{\ben}{\begin{enumerate}}
\newcommand{\een}{\end{enumerate}}
\newcommand{\bit}{\begin{itemize}}
\newcommand{\eit}{\end{itemize}}
\newcommand{\barr}{\begin{array}}
\newcommand{\earr}{\end{array}}
\newcommand{\eps}{\epsilon}

\newcommand{\ber}{\begin{array}}
\newcommand{\eer}{\end{array}}

\begin{document}

\begin{frontmatter}
\title{ A method for enhancing the stability and robustness of explicit schemes in CFD    }

\author{ Hujeirat, A. \thanksref{email}}
\address{Institute of Applied Mathematics, University of Heidelberg, 69120 Heidelberg, Germany }

\thanks[email]{E-mail: ahmad.hujeirat@iwr.uni-heidelberg.de}

\begin{abstract} A method for enhancing the stability and robustness of explicit
schemes in computational fluid dynamics is presented.
The method is based in reformulating explicit schemes in matrix form, which can then be
modified gradually into semi or strongly-implicit schemes.\\
From the point of view of matrix-algebra,  explicit numerical methods
are  special cases in which the global matrix of coefficients is
reduced to the identity matrix $I$. This extreme simplification leads to severe limitation of their
stability range, hence of their robustness.\\
In this paper it is shown that a condition, which is similar to
the Courant-Friedrich-Levy (CFL) condition can be obtained from
the stability requirement of inversion of the coefficient matrix. This condition
is shown to be relax-able, and that  a class of
methods that range from explicit to strongly implicit methods can
be constructed, whose degree of implicitness depends on the number
of coefficients used in constructing the corresponding
coefficient-matrices.  Special attention is given
to a simple and tractable semi-explicit method, which is obtained
by modifying the coefficient matrix from the identity  matrix $I$
into a diagonal-matrix $D$. This method is shown to be
stable, robust  and it can be applied to search for stationary
solutions using large CFL-numbers, though it converges  slower
than  its implicit counterpart. Moreover, the method can
be applied to follow the evolution of strongly time-dependent flows, 
 though it is not as efficient as normal explicit
methods.\\
  In addition, we find that the residual smoothing method accelerates convergence
toward steady state solutions considerably and improves the efficiency of the
solution procedure.

\end{abstract}

\end{frontmatter}
\section{Introduction}

In  the last two decades, tremendous progress has been made
in both computational fluid dynamics (CFD) algorithms and the
computer hardware technologies. The computing speed and memory
capacity of computers have increased exponentially during this
period. Similarly is astrophysical fluid dynamics (AFD), which is
considered nowadays as a  rapidly growing research field, in which modern
numerical methods are extensively used to model the evolution of
rather complicated flows. Unlike CFD, in which implicit methods are
frequently used, the majority of the methods used in AFD are
explicit. Several of them became very popular, e.g., ZEUS (Stone \& Norman 1992), NIRVANA+ (Ziegler 1998), FLASH (Fryxell et al.
 2000), VAC (T{\'o}th et al.1998), THARM (Gammie et al. 2003).
The popularity of explicit
methods arises from their being easy to construct, vectorizable,
parallelizable and even more efficient as long as the dynamical
evolution of compressible flows is concerned. Specifically, for
modeling the dynamical evolution of HD-flows in two and three
dimensions explicit methods are highly superior to-date. For
modelling relativistic flows, Koide and collaborators (e.g., Koide
et al. 2000, 2002, Meier et al. 2001) and Komissarov (2001) have
developed pioneering general relativistic MHD solvers. A rather
complete review of numerical approaches to relativistic fluid
dynamics is given by Mart${\rm\acute{i}}$ \& M\"uller (1999) and
Font (2000). A ZEUS-like scheme for general relativistic MHD has
also been developed and is described in De Villiers \& Hawley
(2003).\\
\begin{figure*}
\begin{center}
{\hspace*{-0.0005cm}
\includegraphics*[width=5.35cm,bb=32 285 579 781,clip]{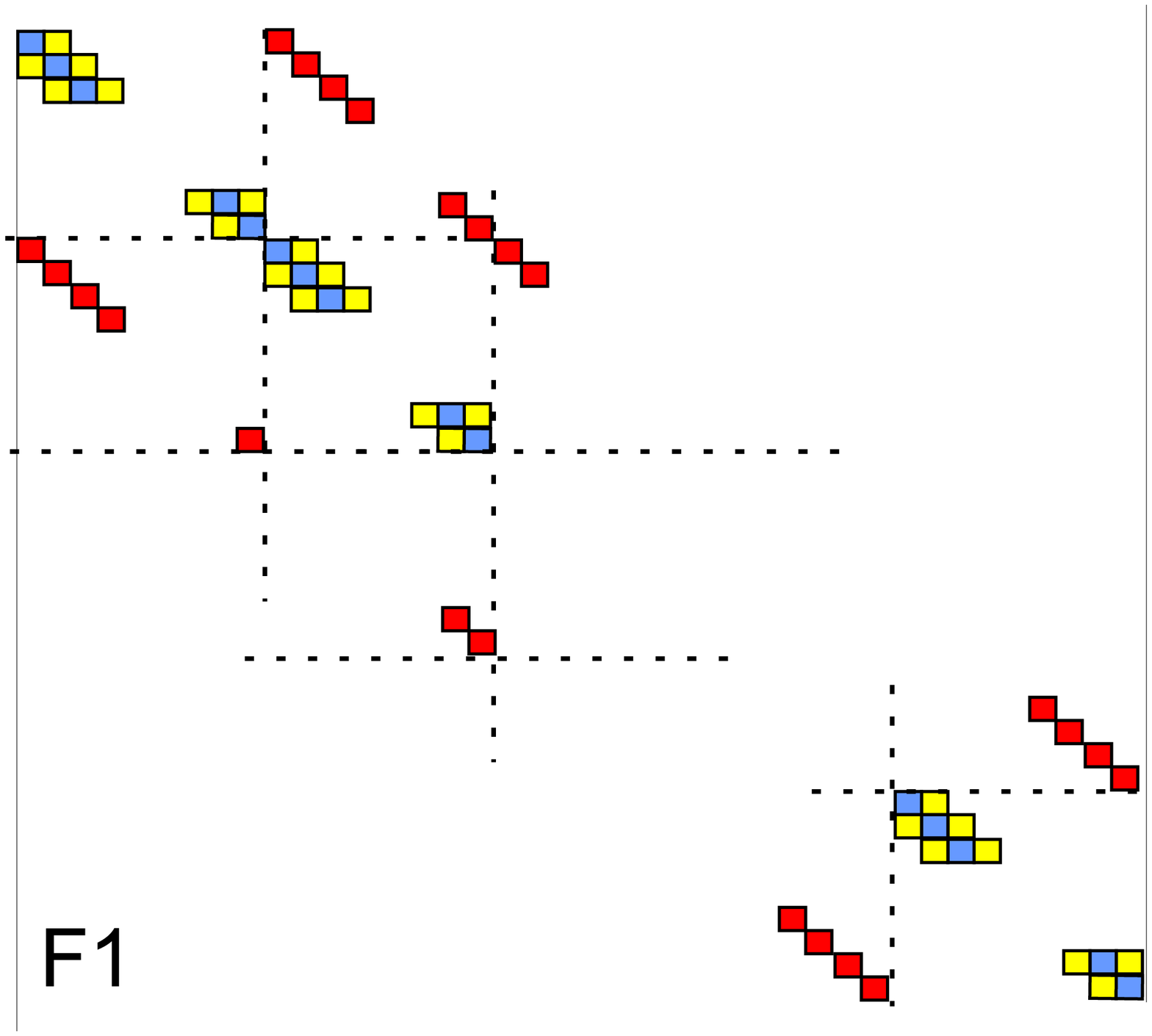}
\includegraphics*[width=5.5cm,bb=57 279 565 743,clip]{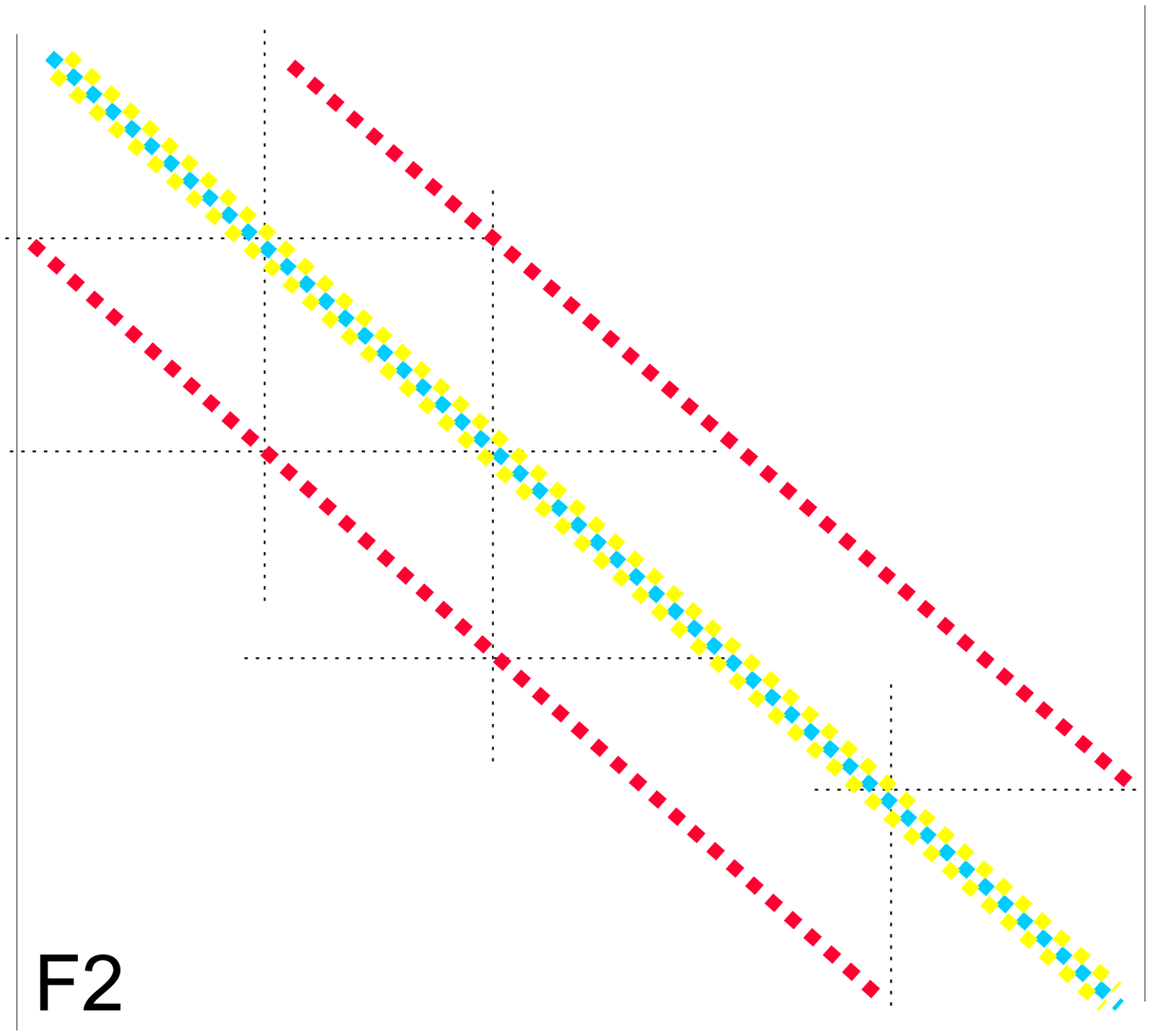}\\
\includegraphics*[width=5.5cm,bb=57 279 565 743,clip]{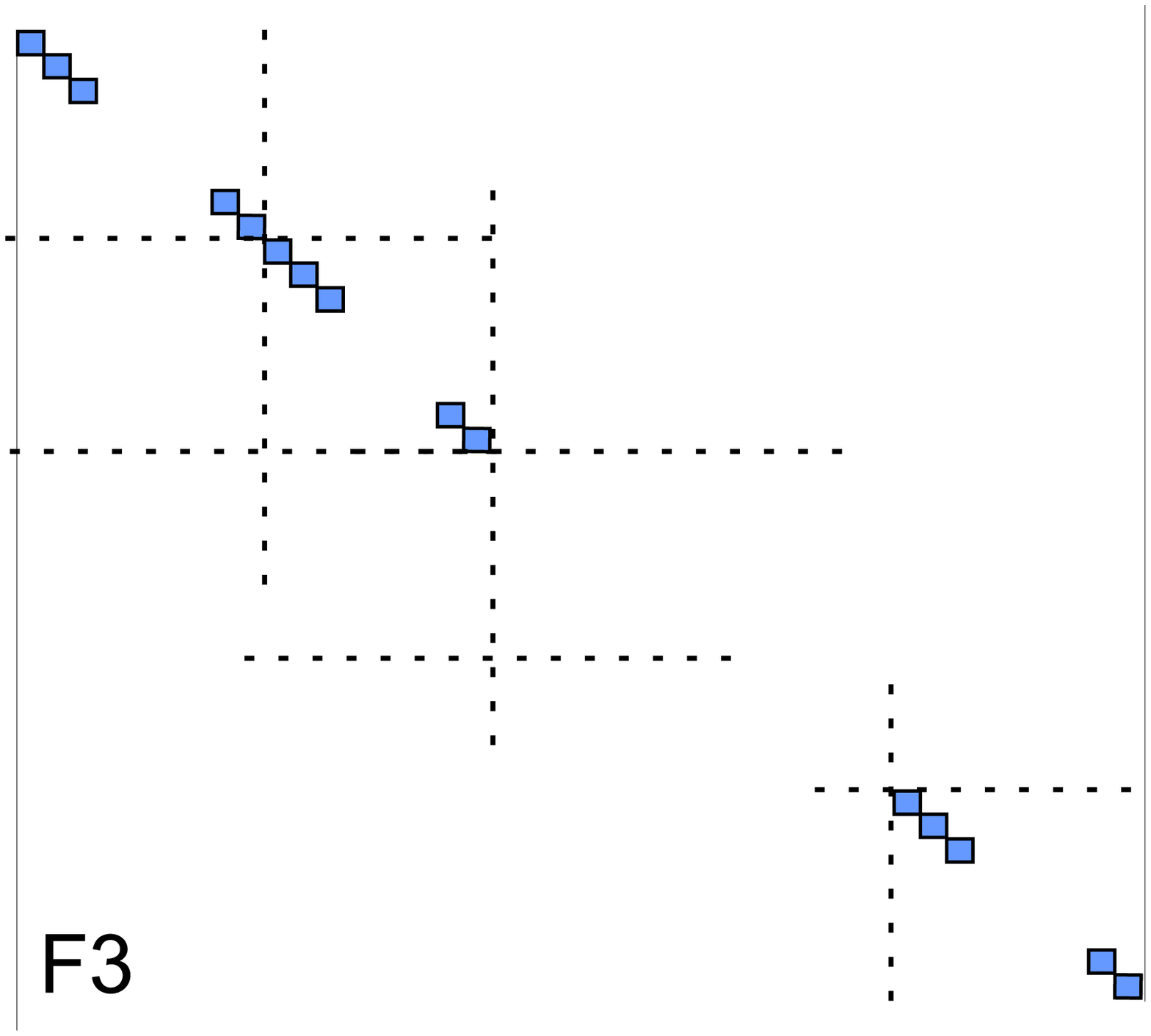}
\includegraphics*[width=5.5cm,bb=57 279 565 743,clip]{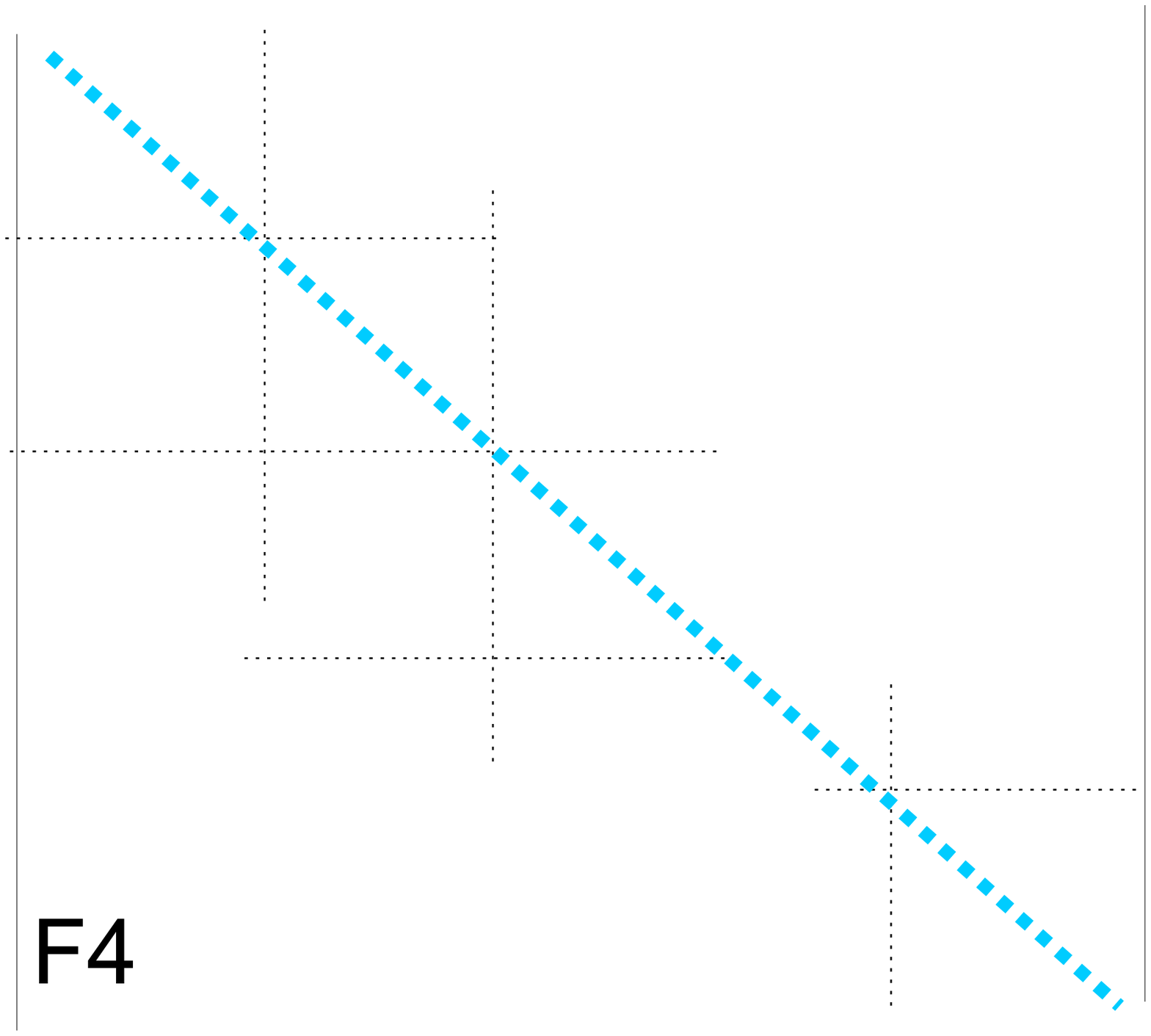}\\
\includegraphics*[width=5.5cm,bb=57 279 565 743,clip]{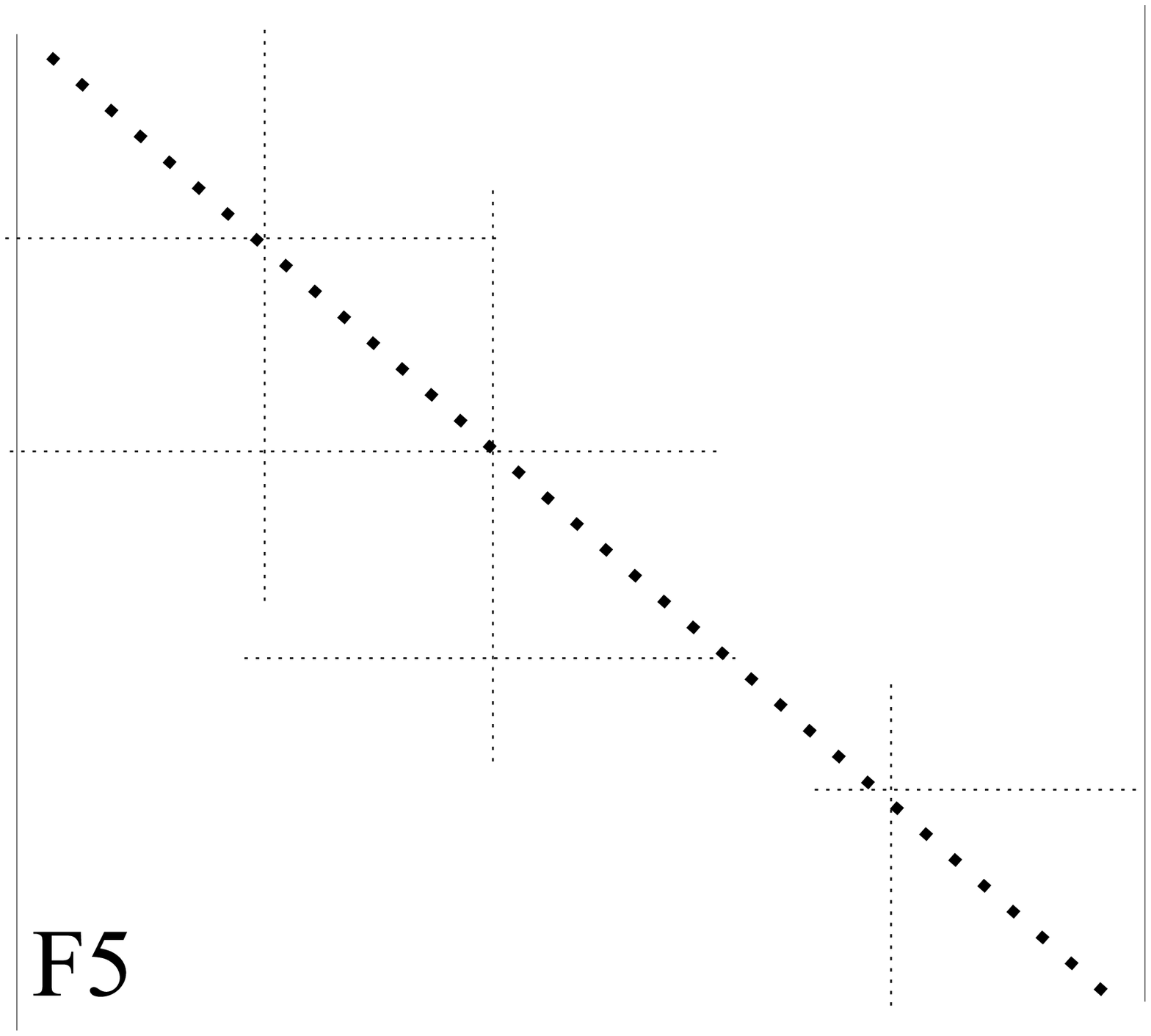} }
\end{center}
{\vspace*{-0.0cm}} \caption [ ] {Schematic description of several coefficient
matrices used by different numerical solution procedures in two
dimensions.  Most commonly used matrices
are F1, F2, F3, F4 and F5, which correspond to
tri-diagonal block, scalar tri-diagonal, block diagonal, scalar diagonal
and  identity matrices, respectively. Solution methods that rely on the
inversion of F1 are classified as implicit, whereas those relying on F5 are explicit.
F2, F3 and F3 correspond to intermediate methods,
i.e., explicit-implicit methods. }
\end{figure*}
However,  these methods are numerically stable as far as the
Courant-Friedrich-Levy number is smaller than unity.  The
corresponding time step size decreases  dramatically with the
incorporation of real astrophysical effects.  Specifically, they
may even stagnate if self-gravity, radiative and chemical effects
are included. Moreover,  explicit methods break down if the flow
is weakly or strongly incompressible, and if the domain of
calculations is subdivided into a strongly stretched mesh.  In an
attempt to enhance their robustness, several alternatives have
been suggested, such as semi-explicit, semi-implicit or even
implicit-explicit methods (Kley 1989, T${\rm\acute{o}}$th et al.
1998 ).
Nevertheless, their rather limited range of applications
has lead to the fact that most  of the interesting
astrophysical  problems remained, indeed, not really solved. A simple
example is the evolution of a steady turbulent accretion disk. It was
found by Balbus \& Hawley (1991) that weak magnetic fields in
accretion disks are amplified, generate turbulence, which in turn
 redistribute the angular momentum in the disk. However, whether this
instability  leads to the long-sought global steady accretion rate, or is it
just a transient phenomenon in which the generation of turbulence is
subsequently suppressed by dynamo action are not at all clear. Other notable phenomena
are the formation and acceleration of the observed superluminal
jets in quasars and in microquasars, the origin of the
quasi-periodic oscillation in low mass X-ray
binaries or the progenitors of gamma ray burst are still spectacular.\\
Explicit methods rely on time-extrapolation procedures for advancing
the solution in time. However, in order to provide physically
consistent solutions, it is necessary that these procedures are
numerically stable.  The usual approach for examining  the
stability of numerical methods is to perform the so called von
Neumann analysis (see Hirsch 1988 for further details). For
example, using an explicit procedure to solve the simple
advection-diffusion equation:
\beq
T_\mathrm{,t}   + u T_\mathrm{,x} = \nu T_\mathrm{,xx} + f,
\eeq
\begin{figure}
\begin{center}
{\hspace*{-0.0005cm}
\includegraphics*[width=6.35cm,bb=14 345 567 803,clip]{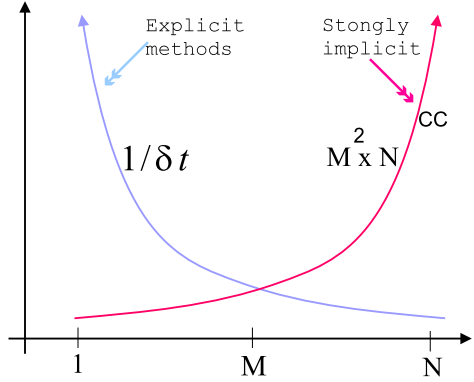}}
\end{center}
{\vspace*{-0.0cm}} \caption [ ] {A schematic description of the
time step size  and the computational costs versus the band width
$M$ of the coefficient matrix. N is the number of unknowns. Explicit methods  correspond to
$M=1$ and large $1/\delta t$. They require minimum computational costs (CC). Large time steps
 (i.e., small $1/\delta t$) can be achieved using strongly implicit methods.
These methods generally rely on the inversion of matrices with large band
width, hence computationally expensive, and, in most cases,
 are inefficient.
  }
\end{figure}
 and applying the von Neumann stability analysis,  it is
necessary, but not sufficient, that the CFL-number is less
than unity. This is equivalent to require
that the time step size fulfills the inequality:
\(
    \delta t \le \min (\DD{\Delta X}{|u| + |\DD{\partial f}{\partial \rho}|^{1/2}}, \DD{\Delta
    X^2}{\nu}),
\)
 where $T,\, \nu,\, f,\, \rho$ denote an arbitrary diffusible variable, viscosity coefficient,
internal or external forces  and density of matter, respectively.
The sub-scripts $t, x, xx$ represent first and second order derivatives of $T$ with respect to
time and to the x-coordinate.
The minimum function in the last inequality runs over each grid
point of the domain  of calculation.
 For additional details see Sec. 9.4 in Fletcher (1991).
 The force $f$ may corresponds
to internal forces,  such as thermal, radiative or magnetic
pressure. In regions where the density is low (e.g., corona around
compact objects such as
 black holes or even above
accretion disks), the derivative $|\DD{\partial f}{\partial \rho}|^{1/2}$
corresponds to the sound speed, speed of light or to the propagational speed of magnetic
{\Alfven} waves, all of which can be extremely large compared to
the actual velocity of the flow.
 Consequently, the CFL-condition limits the range of application and
severely affects the robustness of explicit methods. In particular,
equations corresponding to  physical processes occurring on much
 shorter time scales than the hydro-time scale (e.g., radiation, self-gravitation
and chemical reactions) cannot be followed  explicitly.
Furthermore, these methods are not suited for searching
 solutions that correspond to evolutionary phases occurring on time scales much longer
than the hydro-time scale. Using high performance  computers to
perform a large number of explicit time steps may lead to
accumulation of round-off errors that can easily distort the
propagation of information from the boundaries and cause
divergence of the solution procedure,
 especially if Neumann type conditions are imposed at the boundaries. \\
In contrast to explicit methods, implicit methods are based on solving a matrix
equation of the form $\overline{A}\delta q=d$, where $\overline{A}$  is the coefficient matrix
resulting from the linearization of the system of equations to be solved, $d$ is
the right hand side vector of  known quantities, and $\delta q$ is the solution vector sought.
 These methods have  two
major drawbacks. First, constructing the matrix A is difficult, time
consuming, and  may considerably influence the stability and
robustness of the method. Second, the inversion procedure  must be
stable and extremely efficient. In general, conservative
discretization  of the MHD equations give rise to sparse matrices, or
even to  narrow band matrices. Therefore, any efficient  matrix inversion procedure  must
take the advantage of A being sparse (see Fig. 1). Inverting A directly by
using  Gaussian elimination  requires ${\rm N^{3}}$ algebraic
operations, where N is the number of unknowns (Fig. 2). If the flow is
multi-dimensional and a high spatial resolution is required, the
number of operations can be prohibitive even on modern
supercomputers. Krylov sub-iterative methods (KSIMs), on the
other hand, are most suited for sparse matrices and avoid the
fill-in procedure. In the latter case,  A is not directly involved in the
process, but rather its multiplication with a vector. The
convergence rate of KSIMs has been found to depend strongly on
the proper choice of the pre-conditioner. For advection-dominated
flows, incomplete factorization such as ILU, IC and LQ,
approximate factorization, ADI, line Gauss-Seidel are only a small
sub-set of possible sequential pre-conditioners  (see Saad \& van der
Vorst 2000 and the
 references therein). Another powerful way of accelerating relaxation techniques is to use the
multi-grid method as direct solver or as pre-conditioner (see
Brandt 2001; Trottenberg et al. 2001).
 For parallel computations, Red-Black ordering in combination with GRMES
and Bi-CGSTAB as well as domain decomposition are among the
popular pre-conditioners
(see Dongarra et al. 1998 for further discussion).  \\
In this paper we present a strategy for enhancing the stability
and robustness  of explicit methods.

In Sec.2 of this paper we describe and apply the new strategy
to scalar equations in one-dimension, and generalize it to system of equations in
Sec. 3. Here it is shown that a class of semi-explicit
numerical methods can be easily constructed through reformulating explicit
methods in matrix form, which thereafter can be modified. Spacial attention
is given to a simple semi-explicit method which can be applied for searching 
stationary solutions. Although The method is easy to program and stable even when using $CFL > 1$, it
converges relatively slowly compared to implicit methods,  and therefore further additional 
acceleration techniques of convergence are still to be found.
The results of various test calculations are presented in Sec. 5, followed by
the summary and conclusions in Sec. 6.  
\section{Stability of matrix inversion and  the CFL connection }
In fluid flows, the equation of motion which describes
hydrodynamically the time-evolution of a quantity q in
conservation form reads: \beq
    \DD{\partial q}{\partial t} + L{q} \vec{V} = {f},
\eeq where $\vec{V}$ and $f$ are spatially varying velocity field
and external forces, respectively. $L$ represents  a first and/or
second order linear differential operator that describe the
advection and diffusion of $q$.
In the finite space $\Re$, we may replace the time derivative
of $q$ by: \beq
  \DD{\delta q}{\delta t} =  \DD{q^\mm{n+1} - q^\mm{n}}{\delta t},
\eeq where $q^\mm{n}$ and $q^\mm{n+1}$ correspond to the actual
value of $q$ at the old and new time,
levels, respectively. \\
An explicit formulation of Eq.3 reads: \beq
  \DD{\delta q}{\delta t} = [ -L{q} \vec{V} + {f}]^\mm{n},
\eeq whereas the corresponding implicit form is: \beq
  \DD{\delta q}{\delta t} = [ -L{q} \vec{V} + {f}]^\mm{n+1}.
\eeq Combining these two approaches together, we obtain: \beq
  0= -\DD{\delta q}{\delta t}  + \theta [ -L{q} \vec{V} + {f}]^\mm{n+1}
                          + (1-\theta) [ -L{q} \vec{V} + {f}]^\mm{n} \doteq RHS,
\eeq
where $\theta (0 \le \theta \le 1)$ is a switch on/off parameter.\\
{ The spatial operator ${H} \doteq [ -L{q} \vec{V} + {f}]$ 
should be computed at each grid point, using a conservative
discretization. In this case, the coefficients of the Jacobian matrix can be constructed by 
computing  ${\rm \D RHS}/{\D q^\mm{n+1}} $ which we decompose in   $ A \doteq -[\D { H}/{\D q}]^\mm{n+1}$
 and $I/\delta t$. 
 Thus,  Eq. 6 can be replaced by the matrix-equation:}
 \beq
  (\DD{I}{\delta t}  + \theta A)\delta q =   RHS.
\eeq
As a starting condition in each time step, it is suggested to set $q^{n+1}  = q^{n}.$
Thus, the scheme degenerates into explicit scheme for $\theta=0$ and into iterative implicit for
$\theta =1$.

Eq. 7. is a combination of the Newton and the Crank-Nicolson methods (Fletcher 1991; p. 165 and 304).
It requires that in each time step the RHS should vanish completely. If steady
state or quasi-stationary solutions are sought, then both the RHS as well as  $({q^{n+1}-q^{n}})/{\delta t}$ must
vanish simultaneously. In the latter case, care should be taken to assure that $\theta \rightarrow 1$  
as $\delta t \rightarrow \infty$.\\
We note that since the LHS and the RHS of Eq. 7 are generally different, a loop of iteration within
each time step must be constructed in order to obtain  a reasonable value 
for\footnote{Such deviations may evolve if the LHS of Eq. 7 is not precisely the real Jacobian of Eq. 6, and
if the equations to be solved are partially non-linear .} 
$q^{n+1}$.

\begin{figure}[htb]
\begin{center}
{\hspace*{-0.5cm}
\includegraphics*[width=5.0cm,bb=0 0 310 345,clip]{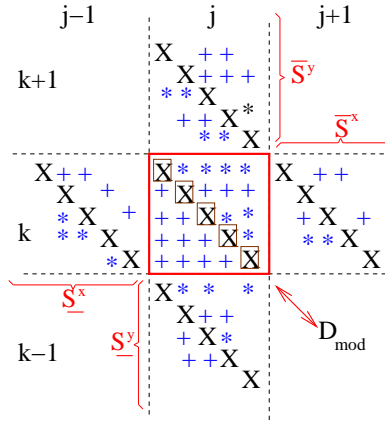}
}
\end{center}
{\vspace*{-0.0cm}} \caption [ ] {The neighboring block matrices in
the x and y-directions resulting
              from 5-star staggered grid  discretization. Entries marked with `X' denote
              the elements usually used in the implicit operator splitting approach
              (Hujeirat \& Rannacher 2001), `*' and `+' are coefficients corresponding
              to the the source terms. The semi-explicit  method for a scalar equation
              relies on inverting the diagonal matrix whose entries are marked
             with X surrounded by squares. The generalization of the
            semi-explicit  method to the multi-dimensional HD-equations requires inverting
            the  block diagonal matrix $D_\mm{mod}$.
  }
\end{figure}
The matrix $A$ contains coefficients such as $V/\Delta x$,
$\eta/\Delta x^2$ that correspond to advection and diffusion
terms, respectively, as well as additional coefficients that
correspond to source terms. \\
{ Noting that Eq. 2 describes the time evolution of a real physical variable $q$, and
applying a conservative discretization to the advection operator $L$, e.g., first
order up-winding, it is easy to verify that the resulting matrix $\bar{A} = I + {\delta t}\theta A$ is
positive definite (nonsingular) and,  diagonally dominant. Therefore,  the  matrix
$\bar{A}$ can be stably inverted (Hackbusch 1994).\\
In this case the relevant question is:    how far can we simplify $\bar{A}$, and still able to provide
corrections $\delta q$,  appropriate for convergence ?\\
Let us replace the real matrix coefficient $\bar{A}$ by a matrix $\tilde{A}$, which is simple and easy to invert.
Thus, instead of solving the ``difficult'' matrix equation  $ \bar{A} \,\delta q = {\delta t}\, RHS$,
we solve $\tilde{A}\, \tilde{\delta q} = {\delta t}\, RHS$. However, in order to assure that
we are dealing with the original problem still,
the matrices $\bar{A}$ and  $\tilde{A}$ must share the essential spectral properties, i.e, the ``spectral
equivalence" (Golub \& Loan 1989; Hackbusch 1994, p. 217). For example in explicit methods  
 $\tilde{A} = I$.
In order that the matrices
$I$ and $\bar{A}$ be spectrally equivalent, 
the sum of the absolute values of the elements in each row of ${\delta t}\,\theta A$  must be smaller
than the corresponding diagonal element of $I$. This implies that
$2\,\theta\,|V|/\Delta x < 1/\delta t$.} The latter condition is relatively strong, and
 a weaker condition can be obtained if the flux difference, rather than the flux itself,  is considered. 
In this case, and 
in the absence of diffusion and external forces, the following inequality holds:
 \beq
  |{\Delta V q}|/\Delta x < |q|/\delta t,
 \eeq
 where $q$ may acquire negative and positive values. However, since $Vq$ is a
conservative
quantity, we may obtain an upper limit for the flux-difference: \\
\beq
  |\Delta V q| \le max\{(Vq)_\mm{in}, (Vq)_\mm{out}\}
  = |V||q|,
\eeq
 where the sub-scripts ${\rm ``in,\, out"}$ denote the locations of the in- and out-flows through
 the surfaces of an arbitrary finite volume cell. In writing the last
 equality, we have omitted these subscripts for simplicity.\\
Thus, in terms of Eq. 7, neglecting $\theta A$ is equivalent to
require:
 \beq
 |{\Delta V q}|/\Delta x \le |V||q|/\Delta x <|q|/\delta t.
 \eeq
This is equivalent to the outcome of the von Neumann stability
analysis (see Richtmyer \& Morton 1967), which yielded the
well-known condition
 $ CFL = |V|\delta t /\Delta x < 1$. \\

It should be noted, however, that the classical derivation of the
CFL conditions relies on the assumption that the velocity V in Eq.
2 is constant. Practically, the HD-equations are non-linear, and
using $CFL =1$ does not prevent the exponential growth of the
instability.  Indeed, most of the  explicit methods used in astrophysical
fluid dynamics
provide stable solutions, if only the $CFL-$number is strictly
less than unity, and in most cases, a  $CFL \le 0.6$ is required.
\begin{figure*}
\begin{center}
{\hspace*{-0.5cm}
\includegraphics*[width=9.5cm,bb=143 180 418 600,clip]{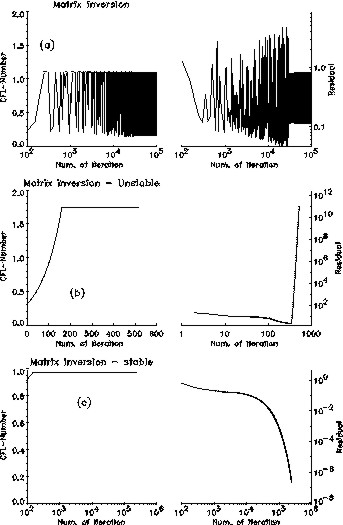}
}
\end{center}
{\vspace*{-0.0cm}} \caption [ ] {
  The evolution of the residual and the CFL-number versus number of iteration
  for the  heat diffusion equation (Eq. 21).
   In (a)  the time step is constructed directly from the residual.
   Here the diagonal elements, or equivalently the time step size,  oscillates around
  a canonical value that corresponds to CFL = 1. The matrix inversion is highly unstable
 and fails to provide a perturbation-free steady solution. In (b) the time step
  is set to increase gradually  from a small value up to
   a value that corresponds to CFL=1.75. The residual in this case grows
exponentially, which implies that the matrix coefficient cannot be
stable-inverted. In (c) the diagonal elements, i.e., $I/\delta t$, are taken to correspond to CFL=0.975. In this case the matrix has a stable inversion, hence
converges, though extremely slow.
     }
\end{figure*}
The matrix $A$ can be decomposed as follows: $ A = D + L + U$,
where D is a matrix that consists of the diagonal elements of $A$.
L and U contain respectively the sub- and super-diagonal entries
of $A$. Noting that a conservative discretization of the
advection-diffusion hydrodynamical equations (Navier-Stokes
equations)
 gives rise to a $D$ which contains positive values, we may
reconstruct a modified diagonal matrix $ D_\mm{mod} = {I}/{\delta
t} + \theta D$. In this case, Eq. 7 is equivalent to:
 \beq
   [D_\mm{mod}   + \theta(L + U)]\delta q = RHS.
 \eeq
A slightly  modified semi-explicit form can be obtained by
neglecting the entries of the matrix $\theta(L + U)$. In this
case,  a necessary condition for the iteration procedure to
converge  is that the absolute value of the sum of elements in
each raw of $\theta(L + U) $ must be much smaller than the
corresponding diagonal element of $D_\mm{mod}$ in the same raw. In
terms of Equation 9,   the method is said to converge if the
entries in each row of $D_\mm{mod}$ fulfill the following
condition:
 \beq 1/{\delta t} + \theta (|V|/\Delta x + \eta/\Delta
x^2 + g ) > ||A-D||_\infty,
 \eeq
 where $||A-D||_\infty$ denotes
the $\infty-$norm of $A-D$, i.e., the maximum row sum of the
modulus of the elements of $A-D$. This condition
 can be fulfilled, however, if the flow is smooth, viscous,  and
 if appropriate  boundary conditions
are imposed\footnote{Note that diffusion pronounces the inequality
in Eq. 12, which gives rise to larger CFL-numbers.}. Consequently,
the inversion process of $D_\mm{mod}\delta q = RHS$
should proceed stable even when large CFL-numbers are used (see Fig. 6 and 7).

As a numerical test, we have applied this approach to the
one-dimensional diffusion equation in spherical geometry (see Eq.
21, see Sec. 5.1,
for a detailed description of the physical problem).\\
 In this particular model, the matrix $A$ of Eq. 7 is removed,
and the equation to be solved reads:
\beq
 (\DD{I}{\delta t}) \delta q = RHS.
\eeq
As a first step, the equation is solved using a direct band matrix solver,
where the band
width is set to 1. Thus, the solver in this case is
exact up to the machine accuracy. The time step size is set to be
determined from the residual, without knowing a-priori about the
CFL associated-problems. The corresponding results are displayed in
Fig. 4. Indeed, we see that the optimal values of the diagonal
elements that limit the exponential growth of
 the residual
 oscillate around a  canonical value which corresponds to CFL=1
 (a/Fig 5). Decreasing the diagonal values artificially,  so to
correspond to CFL number that is  slightly larger than one, gives rise
to an exponential growth of the residual (b/ Fig. 4). On the other hand,
setting the diagonal values to correspond to CFL-numbers that are
slightly smaller than unity, we find that the matrix can be stable-inverted,
and the resulting solutions  converge to the stationary solution, though extremely slowly (c /Fig. 4). \\
To examine the connection between the stability of the
matrix inversion and the CFL condition applied to the heat diffusion
equation in $\Re$,  we may re-write this equation using finite
volume descritization (Fig. 5), and use the defect-correction formulation
of Eq. 7.
\begin{figure}
\begin{center}
{\hspace*{-0.5cm}
\includegraphics*[width=3.5cm,bb=37 43 565 257,clip]{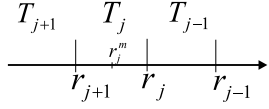}
}
\end{center}
{\vspace*{-0.0cm}} \caption [ ] {In the finite volume discretization, the
temperature $T_j$ is defined at the cell center $r^m_j$, whose boundaries
are $r_j$ $r_{j+1}$.
     }
\end{figure}
Thus, at an arbitrary finite volume cell j,  the equation reads:
 \beq
  [\underline{S}_{j-1}] \delta T_{j-1} + [{1/\delta t} + {D}_{j}] \delta T_{j} + [\overline{S}_{j+1}] \delta T_{j+1}
   = RHS_j,
 \eeq
where $\underline{S}_{j-1} (= -[\nu/vol_j]\times [r^2_j/\Delta
r^m_j])$, $\overline{S}_{j+1} (= -[\nu/vol_j]\times
[r^2_{j+1}/\Delta r^m_{j+1}]),$     and    ${D}_{j} (=
-\underline{S}_{j-1} - \overline{S}_{j+1})$ correspond to the
sub-diagonal, super-diagonal and to the diagonal entries of the
coefficient matrix. Here  $\Delta r^m_j = (r^m_{j-1} - r^m_{j}),\,$
$vol_j = (r^3_j - r^3_{j+1})/3$  and
\(RHS_j = -(\DD{T^{n+1}-T^n}{\delta t})
    + (\DD{\nu}{vol_j}) \{ r^2_j (\DD{T^i_{j-1}-T^i_{j}}{\Delta r^m_{j} })
                         - r^2_{j+1} (\DD{T^i_{j}-T^i_{j+1}}{\Delta r^m_{j+1} })
                         \}^{n+1} + 1.  \)\\

Let us
consider the following three cases:
\ben
 \item The matrix corresponding to Eq. 14 is diagonally dominant for every
   time step size $\delta  t$, hence it can be stable-inverted.
\item If we neglect the sub- and super-diagonal entries and preserve
 ${D}_{j}$, then the
diagonal dominance of the matrix  is not affected, maintaining thereby its
 stability  for any time step size  $\delta t$.
\item If we decide, however, to neglect ${D}_{j},\, \underline{S}_{j-1}$
and  $\overline{S}_{j+1}$, then it is necessary  that $[{1/\delta t}]$ be
larger than each of the neglected values at every grid
point and for all times, i.e., $[{1/\delta t}]
> |\underline{S}_{j-1}| + |\overline{S}_{j+1}|\,\,\, \forall j \in (1,N).$ This is equivalent to require:
 \[\delta t < (\DD{vol_j}{\nu}) (\DD{\Delta r^m_j \Delta r^m_{j+1}}
 {r^2_j \Delta r^m_{j+1} + r^2_{j+1}\Delta r^m_{j}}) .\]
 In plane
 geometry, this inequality  reduces to:
\[\delta t < (\DD{\Delta X_j}{\nu})
 (\DD{\Delta X_j \Delta X_{j+1}} {\Delta X_{j} + \Delta X_{j+1}}),\] which must be fulfilled at every  finite volume
cell. This inequality is  extremely similar to the CFL condition resulting
from von Neumann stability analysis.
\een
  Therefore, the narrow range of stability of explicit methods is a consequence of
 over-simplification of the coefficient matrix.
\section{System of equation - general case}
The set of 2D-hydrodynamical equations in conservative form and
in Cartesian coordinates
 may be written in the following vector form:
\beq
 \DD{\partial \vec{q}}{\partial t} + L_\mm{x,xx} \vec{F} +   L_\mm{y,yy} \vec{G}  = \vec{f},
\eeq
where $F$ and $G$  are fluxes of $q$, and $ L_\mm{x,xx},\,
L_\mm{y,yy}$ are first and second order transport operators
 that describe advection-diffusion
of  the vector variables $\vec{q}$ in x and y directions.
$\vec{f}$ corresponds
to the vector of source functions.\\
By analogy with Eq.7, when  linearization  Eq. 15,  the following
matrix form can be obtained: \beq [\DD{I}{\delta t} +
\theta(AL_\mm{x,xx} + B L_\mm{y,yy}-H)]\delta q = RHS, \eeq
where $ A [=\partial F/\partial q],\,  B[=\partial G/\partial q]$ and $ H[= \partial \vec{f}/\partial
q]$, are evaluated on the former (or the new, if Newton method is
used to construct the matrices) time levels.
$RHS^\mm{n}=[\vec{f}- L_\mm{x,xx} \vec{F} -   L_\mm{y,yy} \vec{G}]^\mm{n}$.\\
Adopting a five star staggered grid discretization, it is easy to
verify that at each grid point Eq. 16 acquires the following block
matrix equation:
\[
\DD{{\delta q}_\mm{j,k}}{\delta t} + \underline{S}^\mm{x}{\delta
q}_\mm{j-1,k} + {D}^\mm{x}{\delta q}_\mm{j,k}
                + \overline{S}^\mm{x}{\delta q}_\mm{j+1,k} \]
\beq
 + \underline{S}^\mm{y}{\delta q}_\mm{j,k-1} + {D}^\mm{y}{\delta q}_\mm{j,k}
                + \overline{S}^\mm{y}{\delta q}_\mm{j,k+1}
= RHS_\mm{j,k}, \eeq where the underlines (overlines) mark
the sub-diagonal (super-diagonal) block matrices  in the
corresponding directions (see Fig. 3).
 ${D}^\mm{x,y}$ are the
diagonal block matrices resulting from the discretization of the
operators
$L_\mm{x,xx}\vec{F}$,  $L_\mm{y,yy} \vec{G}$ and $\vec{f}$.\\
To outline the directional dependence of the block matrices, we
re-write Eq. 17 in a more compact form: \beq
 \begin{array}{lll}
 & {\hspace*{0.3cm}}\overline{S}^\mm{y}{\delta q}_\mm{j,k+1} & \\
+ \underline{S}^\mm{x}{\delta q}_\mm{j-1,k} & +
{D}_\mm{mod}{\delta q}_\mm{j,k} &
  + \overline{S}^\mm{x}{\delta q}_\mm{j+1,k}   = RHS_\mm{j,k} \\
& + \underline{S}^\mm{y}{\delta q}_\mm{j,k-1}. &
\end{array}
\eeq
\begin{figure}
\begin{center}
{\hspace*{-0.5cm}
\includegraphics*[width=6.5cm,bb=217 183 355 600,clip]{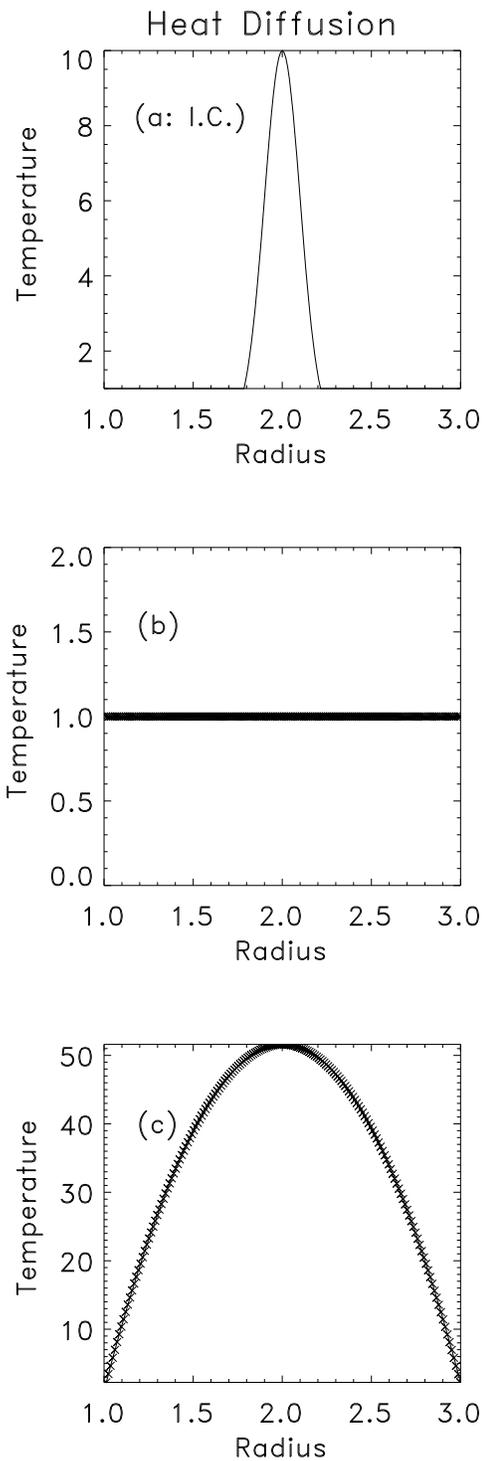}
}
\end{center}
{\vspace*{-0.0cm}} \caption [ ] {  The one-dimensional heat diffusion problem. The profile in (a)
 corresponds to the initial distribution of the temperature. The T-profile in (b) is a
steady solution which was obtained without a source term, i.e., by solving the
equation:
    \(T_{,t}  = r^{-2} (r^2 \nu T_{,r})_{,r} \)
(see Sec. 5.1).
The profile (c) corresponds to the stationary solution obtained with a
source term (see Eq. 21).
  }
\end{figure}
\begin{figure}
\begin{center}
{\hspace*{-0.5cm}
\includegraphics*[width=6.5cm,bb=200 183 380 600,clip]{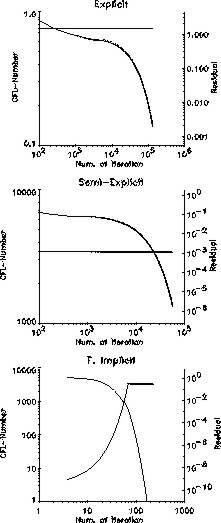}
}
\end{center}
{\vspace*{-0.0cm}} \caption [ ] { The one-dimensional heat diffusion problem.
The evolution of the CFL-number ($\nu\,\delta t/\Delta x^2,$ solid line) and the residual
(dashed line, $RHS$ of Eq. 7) versus number of iteration for different numerical methods. While the
explicit method stagnates after 130000 time steps, the semi-explicit method
converges to the steady solution after 20000 time steps. On the other hand, the
fully implicit method provides the sought solution  after
100 iterations only. Although in the latter case the defect-correction
strategy in combination with several global iterations have been employed to assure
convergence, the implicit solution procedure is remarkably more robust and
more efficient than the former methods.  }
\end{figure}
  where \({D}_\mm{mod} = {{\delta q}_\mm{j,k}}/{\delta t} +
{D}^\mm{x}+ {D}^\mm{y}.\) The subscripts ``j'' and ``k'' denote
the grid-numbering in the x and y directions, respectively
 (see Fig. 3).
Eq. 18 gives rise to three different types of solution procedures:
\ben
\item Classical explicit methods are obviously very special cases that are recovered
      when the sub- and super-diagonal block matrices together with
         ${D}^\mm{x}$ and ${D}^\mm{y}$ are neglected. The only  matrix to be retained here
        is $ (1/{\delta t})\,\times\,$(the identity matrix), i.e.,
        the first term on the LHS of Eq. 16. This yields the vector equation:
    \beq
        \DD{I}{\delta t}{\delta q}_\mm{j,k}  = RHS_\mm{j,k}.
    \eeq
\item Semi-explicit methods are recovered when neglecting the sub- and super-diagonal
  block matrices only, but retaining the block diagonal matrices. In this case
      the matrix equation reads:
    \beq
       {D}_\mm{mod}{\delta q}_\mm{j,k} = RHS_\mm{j,k}.
    \eeq
       We note that inverting ${D}_\mm{mod}$ is a straightforward procedure, which can be maintained
       analytically or numerically.
\item A fully implicit solution procedure requires retaining
       all the  block matrices on the LHS of Eq. 18. This yields a global matrix that
       is highly sparse (a/Fig.1).  In this case, the
      ``Approximate Factorization Method'' (-AFM: Beam \& Warming 1978) and the
     ``Line Gauss-Seidel Relaxation Method'' (-LGS: MacCormack 1985)
       are considered to be efficient solvers for such a set of HD-equations
        in multi-dimensions.
\een 
\subsection{ Residual smoothing and accelerating convergence}
     Let [a,b] be the interval on which Eq. 2 should to be solved. We may divide [a,b] into
    N equally spaced finite volume cells: $\Delta x_\mm{i} = (b-a)/N$, $i=1,\, N$. To follow
    the time-evolution of q using a classical explicit method, the time step size
    must fulfill  the CFL-condition, which requires ${\delta t}$ to be smaller than the
    critical value:
    $ {\delta t}^\mm{u}_\mm{c} $
    If [a,b] is divided into N highly stretched finite volume cells,
    for example $ \Delta x_\mm{1} < \Delta x_\mm{2} ... < \Delta x_\mm{N}$, then the CFL-condition
    restricts the time step size to be even smaller than
    $ {\delta t}^\mm{nu}_\mm{c} = {\min_{i}}\{(\Delta x_\mm{i}/(V + V_\mm{S})_\mm{i}\}$, which
    is much smaller than ${\delta t}^\mm{u}_\mm{c}$.  Thus,
    applying a conditionally stable method to model flows, while using a highly non-uniform
    distributed mesh, has the disadvantage that the time evolution of the variables in the whole domain
    are artificially and severely affected by the flow behaviour on the finest cell.\\
    Moreover, time-advancing the variables may stagnate if the flow is strongly or nearly
    incompressible.
    In this case, $V_\mm{S} >>  V,$ which implies that the time step size allowed by the CFL-condition
    approaches zero. \\
    However, we may  still associate  a time step size with each grid point, e.g.,
    $ {\delta t}^\mm{nu}_\mm{i} = \Delta x_\mm{i}/(V + V_\mm{S})$,  and follow the time
    evolution of variable $q$ in each cell independently. Interactions between variables
     enter the solution procedure through the evaluation of the
     spatial operators on the former time level.
    This method, which is occasionally  called the ``Residual Smoothing Method'' proved to be efficient
    at providing quasi-stationary solutions within a reasonable number of iterations, when compared to
    normal explicit methods (Fig. 10, 13, also see Swanson \& Turkel 1997; Enander 1997).

\begin{figure*}
\begin{center}
{\hspace*{-0.5cm}
\includegraphics*[width=13.5cm,bb=160 150 530 615,clip]{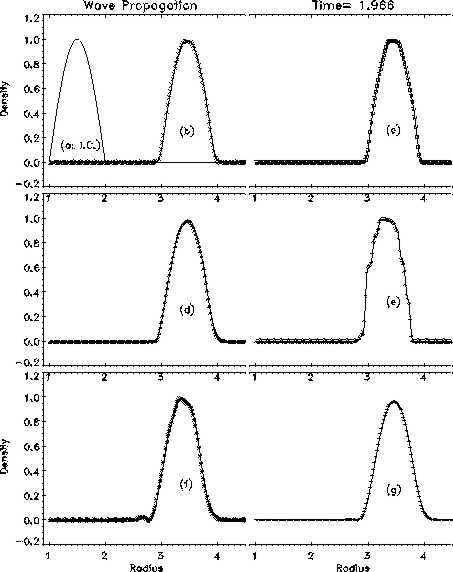}
}
\end{center}
{\vspace*{-0.0cm}} \caption [ ] { The wave propagation problem.
The initial density profile is the sinus wave  (a), which moves
right-wards with velocity $U=1$.  The computed profiles ``b" to
``g" have been obtained after time = 1.966, using 200 finite
volume cells. The adopted advection scheme is of third order
spatial accuracy and second order accurate in time. The profiles
``b, d, f" have been obtained using the fully implicit solution
procedure with time step sizes that correspond to CFL=1, 2.5 and
5, respectively. The profiles ``c, e, g" have been obtained using
the semi-explicit approach with CFL=1, 2.5 and 5, respectively.
  }
\end{figure*}
The main disadvantage of this method is its inability to provide
physically meaningful time scales for features that possess
quasi-stationary behaviour. Here we suggest to use the obtained
quasi-stationary solutions as initial configuration and re-start
the calculations using a uniform and physically relevant time step
size.
\begin{figure}[t]
\begin{center}
{\hspace*{-0.5cm}
\includegraphics*[width=7.5cm,bb=50 273 276 740,clip]{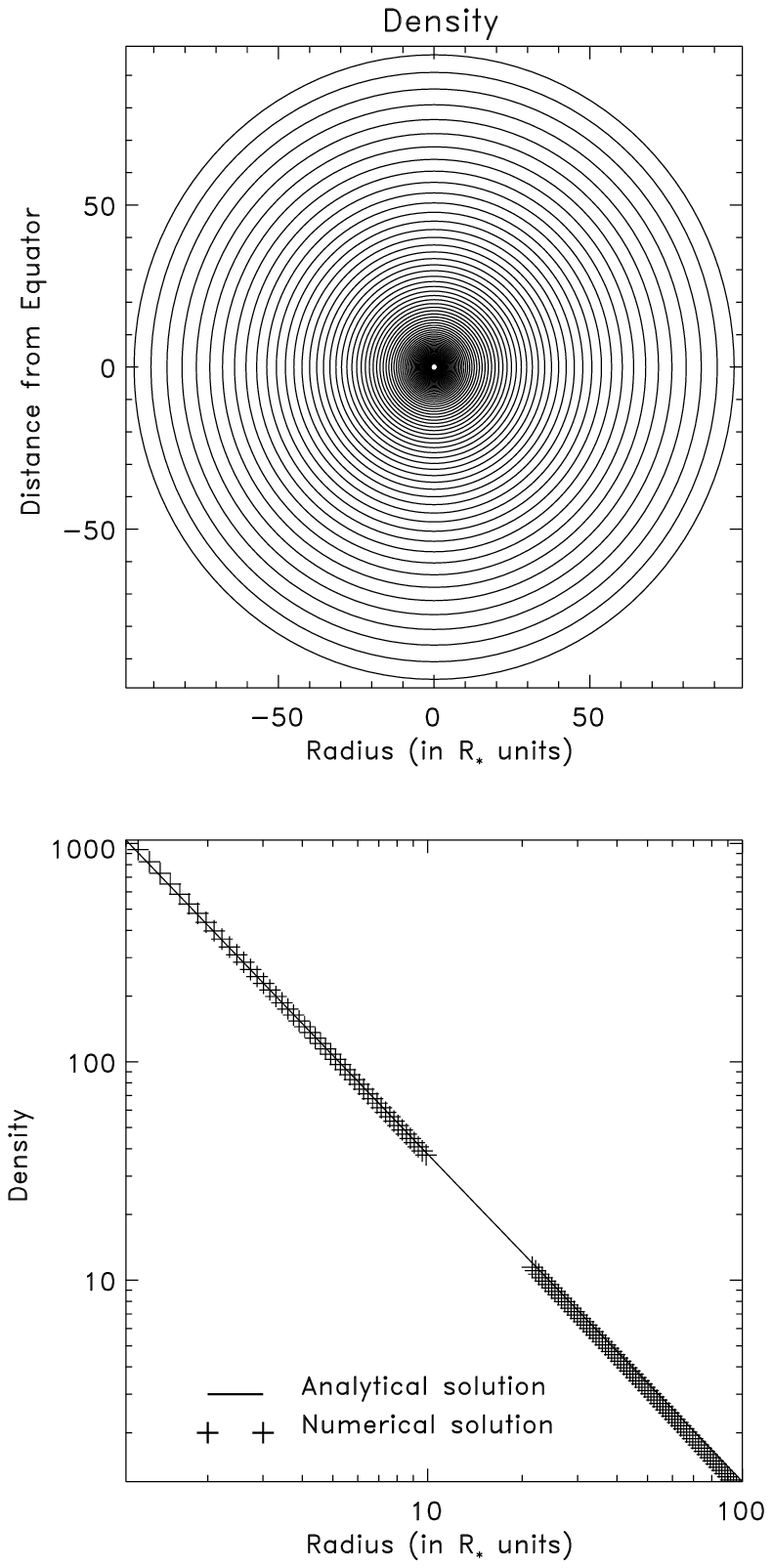}
}
\end{center}
{\vspace*{-0.0cm}} \caption [ ] { Free-fall of gas onto a
non-rotating black hole. Top: 50 density contours of a
freely-falling gas around  one solar-mass black hole. Bottom: the
numerically-obtained density distribution has a power law that
 coincides precisely with the theoretical solution $r^{-3/2}$. In
this calculation, inflow (outflow) at the outer (inner) boundary
have been imposed. 200 strongly-stretched finite volume cells in
the radial direction and 60 in horizontal direction have been
used.
  }
\end{figure}
\section{Other similar approaches}
The Dufort-Frankel scheme is most suited for modelling
diffusion-dominated flows and provides solutions of second order
spatial and temporal accuracies (see Fletcher 1991 for further
details). The scheme is unconditionally stable, as it relies on
adding the positive coefficient resulting from finite difference
discretization of the diffusion operator to the diagonal elements,
thereby enhancing
the diagonal dominance of the matrix coefficient.\\
Despite this similarity, the scheme differs still from the
explicit-implicit methods presented here at least in  two aspects:
 \ben
\item The DuFort-Frankel scheme is generically second order accurate in time. The
      scheme fails to be consistent when a large time steps are used. Thus, the method
      is not suited for searching steady state solutions.
\item The scheme is not suited for modelling advection-dominated flows.
\een
Another solution procedure that might look  similar to the
semi-explicit method presented here is the Jacobi iteration
method. Inspection of Eq. 20, however, shows that neither the
lower nor the upper diagonal matrices are considered in the
present solution scheme. Furthermore, the consistency of our
solution procedure with the original set of equations is
guaranteed through adopting the defect-correction strategy. Unlike
Jacobi method which may diverge if large time steps are used, the
test calculations presented here show that the semi-explicit
method converges even for relatively large CFL numbers.\\
Recently, Yabe et al. (2001) have presented the  Constrained Interpolation Profile (CIP)
 method, which has been successfully applied to model solid, liquid, gas and plasmas.
The method is a kind of semi-Lagrangian scheme which can re-produce strongly
time-dependent solutions with large CFL-numbers. Therefore, the method is promising
and testing its robustness  and capability at modelling astrophysical phenomena
is extremely useful.
\begin{figure*}
\begin{center}
{\hspace*{-0.5cm}
\includegraphics*[width=12.5cm,bb=40 175 385 600,clip]{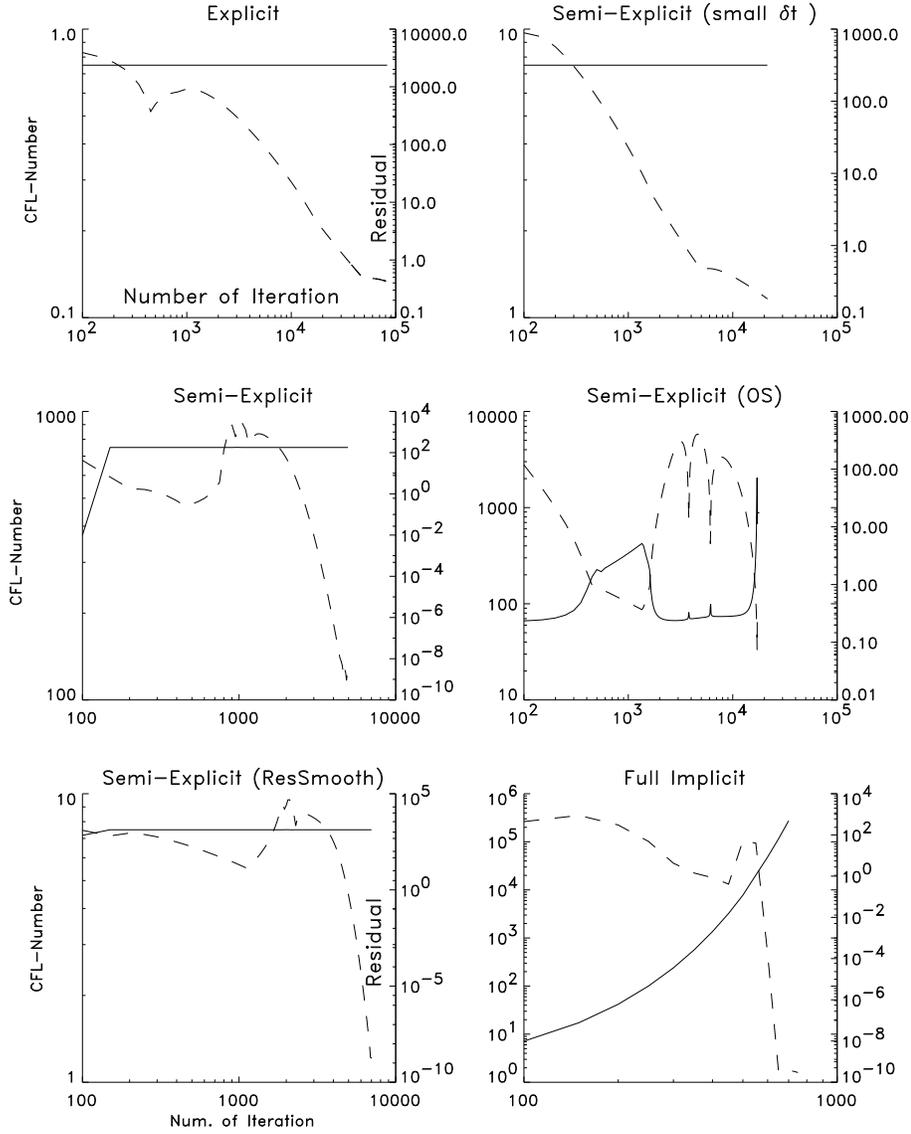}
}
\end{center}
{\vspace*{-0.0cm}} \caption [ ] { Plasma in Free-fall onto a black
hole. Similar to Fig. 7, the evolution of the CFL-number and the
residual versus number of iteration are shown. The solution methods
used here are: normal explicit (top/left), semi-explicit
(middle/left), semi-explicit in combination with the residual
smoothing strategy(bottom/left), semi-explicit using moderate
CFL-numbers (top/right), semi-explicit method in which the time
step size is taken to be  a function of the maximum residual
(middle/right), and finally the fully implicit method
(bottom/right). Obviously, the different forms of the
semi-explicit method used here are stable and converges to the
stationary solution, though at different rates.
  }
\end{figure*}
\section{Test calculations}
In the following, we present the results of several test
calculations aimed to examine the stability, robustness and the
capability of the semi-explicit solver to treat strongly
time-dependent or steady inviscid flows governed by strong shocks.
\subsection{ The diffusion equation }
Diffusion of heat is generally expressed by a second order differential operator.
Taking into account this operator in the von Neumann stability analysis, it can be shown
that it imposes a further limitation on the time step size, hence narrowing
the stability regime of explicit methods even further. Since this limitation
may easily lead to complete stagnation of these methods, especially when steady state
solutions are sought, several modification strategies have been suggested.
A popular strategy is to treat the diffusion operators implicitly.
In the absence of advection and other source terms, however, this modification is
significant, as the scheme is then fully implicit, and therefore have the
usual drawbacks of implicit methods.

The semi-explicit method presented here is almost as efficient as
normal explicit methods. The main difference between these two approaches
is that semi-explicit methods require additional programming efforts.
Specifically, the positive entries resulting from finite differencing
of the operators involved should be carefully selected and added to
the diagonal elements of the coefficient matrix.

The purpose here is to apply the semi-explicit method to the heat diffusion
problem and compare its convergence behaviour with those of explicit and fully
implicit methods.  \\
{ Similar to the other test calculations presented in this paper, 
the solvers here are constructed using routines from the solver package ``IRMHD"
(Hujeirat \& Rannacher 2001). The majority of these routines are
designed and programmed in spherical geometry. Therefore,  to use them properly,
 the equations of interest should be reformulated in spherical geometry.
 Unlike in plane geometry, the main drawback of this procedure are the difficulties
associated with  constructing relevant analytical solutions in spherical geometry. 
On the other hand, locating the domain of calculation far from the
center, may reduce the difference between these geometries ($\sim 1/r$), and which can be made
negligibly small.  }

Thus, the one-dimensional diffusion equation in spherical geometry reads:
\begin{equation}
    \DD{\partial T }{\partial t} = \DD{1}{r^2} \DD{\partial}{\partial r} r^2 \nu
    \DD{\partial T}{\partial r} + 1,
\end{equation}
where $T,\,\nu$ denote the variable temperature and the constant diffusion coefficient, respectively.
This equation is solved in the radial direction in the
interval $[r_\mathrm{in}, r_\mathrm{out}] =[1000,1003],$ which is sub-divided into 180 finite volume  cells
of equal size. The heat equation is solved using the boundary conditions:
$T(r_\mathrm{in})= T(r_\mathrm{out}) = 1.$
Note that the inner and outer boundaries are chosen to be
very close  in order to reduce  geometric compression.
Furthermore, to pronounce the difference in the convergence history between
the explicit, semi-explicit and fully implicit methods, we set
 $\nu=10^{-2}$  and use an initial configuration
that is very different from the sought steady solution. Specifically,
the calculations have been initialized using the starting profile:
$ T(r,t=0) = 10\, e^{-10(r-r_0)^2},$ where
$r_o= (r_\mathrm{in}+r_\mathrm{out})/2$ (see Fig.12).\\
 To first order in (1/r), the steady solution is very close to:
$ T(r, t=\infty) = -\DD{1}{2 \nu} (r-r_\mathrm{in})(r-r_\mathrm{in}) + 1.$

{ In Fig. 6 and 7 the results of three different solution procedures are presented.
For simplicity, the T-profiles in this figure are plotted versus $\bar{r} = r -1000$.}
Obviously, the semi-explicit method converges to the sought steady solution  much
faster than explicit, but extremely slower than the fully implicit solver.
Moreover, the method is numerically stable even when extremely large CFL-numbers
are used.
\subsection{ Wave propagation in one dimension}
In an attempt to examine the capability of the method at capturing
strongly time dependent features, the following one-dimensional
wave equation in spherical geometry has been solved:
\begin{equation}
    \DD{\partial \rho }{\partial t} + \DD{1}{r^2} \DD{\partial}{\partial r} r^2 \rho U =
    0,
\end{equation}
where $\rho, U$  denote  the density and radial velocity,
respectively. The velocity assumes the constant value $ U = 1$. To
reduce the effect of geometrical compression, the interval of
calculation is taken to be $[100 \le r \le 104]$. Similar to the
previous test problems, different solution procedures have been
adopted using advection
schemes of third order spatial accuracy and second order accurate in time.\\
The initial density distribution, is taken to be a  sinus profile
that moves right-wards with velocity $U=1$ ( see Fig. 8, plot a).
As in plane geometry, the shape of the initial density profile is
expected to be preserved as it moves right-wards.\\
To compare results, a reference solution has been produced using
an explicit solution procedure.  The interval of calculation has
been sub-divided into 1000 finite volume cells, and a time step
size that corresponds to CFL = 0.01 has been used. The advection
scheme here is third order accurate in space and second order in
time.

Obviously, the semi-explicit method is capable at capturing the
propagation of the wave front accurately (see Fig. 8). The
amplitude of the wave is hardly changed, even when using a time
step size that corresponds to a CFL-number larger than unity.
However, in the latter case, several iterations per time step were
required to assure convergence. For example, the profile ``e" of
Fig. 8 has been obtained using CFL= 2.5 in combination with two
global iterations per time step, which apparently are not
sufficient for convergence.   However, increasing the number of
iteration from 2 to 7, a more accurate solutions has been obtained
even when the CFL number is increased from 2.5 to 5 (see Fig. 8
/profile ``g"). { Consequently, convergence can be maintained if a sufficient
number of global iterations is 
performed\footnote{Using high spatial and temporal accuracies may lead to significant
deviations of the LHS from the RHS of Eq. 7. This has the consequence
that several iterations might be still required to reduce the RHS to below
a certain critical value.}, though the exact 
correlation is hard to deduce. We note,
however, that the LHS and RHS of Equations 7 or 16 may differ
significantly from each other if the flow is strongly
time-dependent. Here the RHS is calculated using higher order
spatial and temporal accuracies, whereas the LHS is calculated
using first order accuracies. Maintaining compatibility, in this
case, depends on the number of global iterations performed, which
is expected to increase with increasing the CFL-number, but which may diverge
if the flow contains discontinuities or shocks.}

The profile ``f" of Fig. 8 show that even fully implicit methods
(without iteration, hence the LHS and RHS are weak compatibile)
may fail to reproduce the correct wave profile, if large
CFL-numbers are used.

According  to   Eq. 6, using $\theta < 1$ yields a certain
average of the $q-$values on the new and old time levels. This
averaging-process is useful if the sought solutions are strongly
time-dependent, such as propagations of shock waves, where using
large time-steps may cause divergence of the solution track, or
even numerically-destabilize the scheme. A possible way to capture
both time-dependent and quasi-stationary solutions is to relate
$\theta$ to the time step size, which in turns depends on the
total residual. An example is the damped Crank-Nicolson scheme:
 $\theta = (1+\alpha~\delta t)/2 \le 1,$ where $\alpha$ is a constant of
 order unity (see Hujeirat \& Rannacher 2001 and the references therein).
In the case that stationary solutions are of interest, $\theta =
1$ should be used.\\

\subsection{Free-fall  onto a  black hole}
A non-rotating gas around a spinless black hole is gravitationally bound, and
therefore should fall-freely onto the black hole, provided that no
other forces oppose gravity.
The density and
velocity profiles of a freely-falling matter in the radial
direction and far from the event horizon obey the power laws:
$r^{-3/2}$ and $r^{-1/2}$, respectively. This physical problem is
relevant for testing the capability of the numerical approach at
capturing steady and oscillation-free solutions, even when
a strongly stretched mesh distribution is used.\\
The equations to be solved in this problem are the continuity, the
radial and horizontal momentum equations, and the internal energy
equation. The flow is inviscid and adiabatic with the adiabatic
index  $\gamma =5/3$. The equations have been solved using a
first order accurate advection scheme both in space and time.   In
carrying out these calculations, the following conditions/inputs have
been taken into account.
\begin{itemize}
    \item The central object is   a one  solar-mass
    non-rotating black hole.
    \item { The outer boundary is  100 times larger than the
    the inner radius, i.e.,  $R_\mathrm{out}= 100\times
    R_\mathrm{in}$, where $R_\mathrm{in}$ is taken to be the radius of the last stable
    orbit\footnote{$R_\mathrm{LS} = 3\times R_\mathrm{S} = 6\times R_\mathrm{g}$, where $R_\mathrm{S}$ and
     $R_\mathrm{g}$ are the Schwarzschild and 
    gravitational radii, respectively.} $R_\mathrm{LS}$ . To first order in $V/c$, the flow at this radius  
    can be still treated as non-relativistic, though the error can be as large as 30\%.    }
    \item Along the outer boundary,  the density and temperature of the gas
     assume uniform distributions, and flow across
     this boundary with the free-fall velocity. Symmetry boundary
    conditions along the equator, and asymmetry boundary conditions along the
     axis of rotation have been imposed. Along the inner boundary,
     we have imposed non-reflecting and outflow conditions.
     { This means that  up-stream conditions are imposed, which forbid
     information exterior to the boundary to penetrate into the domain of calculations.
     In particular, the actual values of the density, temperature and momentum in the ghost zone  r
     are erased and replaced by the corresponding values
     in the last zone, i.e, the zone between $R_\mathrm{in}$ and $R_\mathrm{in} + \Delta R$.
     In the case that second order viscous operators are considered, care has been taken to
     assure that their first order derivatives across $R_\mathrm{in}$ are vanished.  }
\end{itemize}
The above set of equations are solved in the first quadrant $[1\le
r \le 100]\times [0\le \theta \le \pi/2]$, where 200 strongly
stretched finite volume cells in the radial direction and 60 in
the horizontal direction
are used.\\
Fig. 9 shows the 2D distribution of the density around the hole.
The bottom plot of Fig. 9 shows the precise agreement of the
numerical with the theoretical solution. \\
As in  Fig. 7, we show  in Fig. 10 the evolutions of the CFL-number
and the residual as function of the number of iteration which has
been obtained using different numerical approaches. The results
show that the convergence of the explicit and  semi-explicit
methods are rather slow when a relatively small time step size is
used. This implies that  the amplitude-limited  oscillations are
strongly time-dependent that may result from geometric
compression. Indeed, these perturbations disappear, when
relatively large
time-step sizes are used (see Fig. 10,  middle/right). \\
In addition, the semi-explicit solver has been tested in
combination with the residual smoothing strategy. As expected,
this approach   accelerates  convergence considerably ( Fig. 10:
compare the plots bottom/left with  the top/right).

In most of the cases considered here, the time-step size is set to
increase in a well-prescribed manner and independent of the
residual.  However, determining  the size of the time step  from
the residual directly did not provide
satisfactory convergence histories (Fig. 10, middle/right). \\
The results obtained here indicate that the semi-explicit method
is stable when compared to normal explicit methods. Furthermore,
the semi-explicit method can be applied to search for stationary
solutions using large time steps, or equivalently,  CFL-numbers
that are significantly larger than unity (Fig. 10, middle/left).
\subsection{Shock formation around black holes}
Similar to the forward facing step in CFD,
a cold and dense disk has been placed in
the innermost equatorial region: $[1\le r\le 10]\times [-0.3 \le
\theta \le 0.3]\footnotemark[1]$.
\footnotetext[1]{ In applying spherical geometry, the transformation
$\bar{\theta} = \pi/2 - \theta $ has been used. This is useful if
further transformation into cylindrical geometry is planned.}
We use the same parameters, initial and boundary condition
use in the previous flow problem. A vanishing in- and out-flow conditions have
been imposed at the boundaries of this disk. The gas
surrounding the disk is taken to be inviscid, thin, hot and
non-rotating. Thus, the flow configuration is similar to the
forward facing step problem usually used for test calculations in
CFD. The disk here serves as a barrier that forbids the gas from
freely falling onto the black hole, and instead, it forms a curved
shock front around the cold disk. The purpose of this test is
mainly to examine the capability of the method at capturing steady
solution governed by strong shocks.
\begin{figure}
\begin{center}
{\hspace*{-0.5cm}
\includegraphics*[width=6.25cm,bb=45 33 275 740,clip]{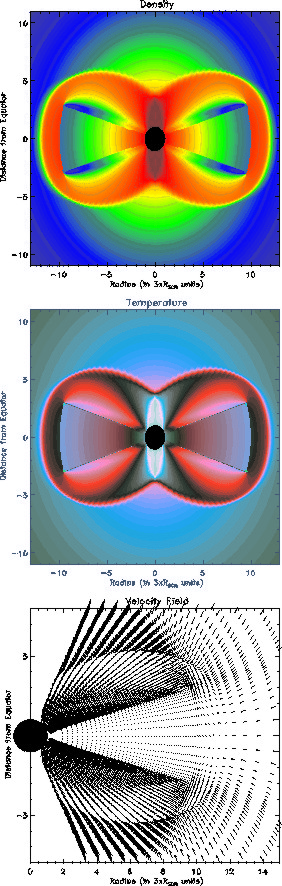}
}
\end{center}
{\vspace*{-0.0cm}} \caption [ ] {  Free-fall of gas onto a black
hole surrounded by a cold disk. Top: the density distribution
 (red: large values, blue: low values, green: intermediate values).
  Middle: the temperature distribution (red: large values, blue: low
  values, gray: intermediate values). The curved shock front,
  where the temperature attains maxima is obvious.
  Bottom: the distribution of velocity field.
  }
\end{figure}
 In solving
the HD-equations (see Sec. 4.1), an advection scheme  of third
order spatial accuracy and first order accurate in time has been
used. The domain of calculation is sub-divided into 200
strongly-stretched finite volume cells in the radial direction and
60 in horizontal direction. In Fig. 11 the configuration of the
steady distributions of the density, temperature and the velocity
field are shown. Similar to the calculation in the previous
sections, the results indicate that the semi-explicit method is
stable and converges to the sought steady solution even when a
CFL-number of order 200 is used (see Fig. 12). However,
the method converges relatively slowly compared to the implicit
operator splitting approach, where steady solutions have been
obtained after one thousand iterations only.
\begin{figure}[htb]
\begin{center}
{\hspace*{-0.5cm}
\includegraphics*[width=7.5cm,bb=40 317 215 602,clip]{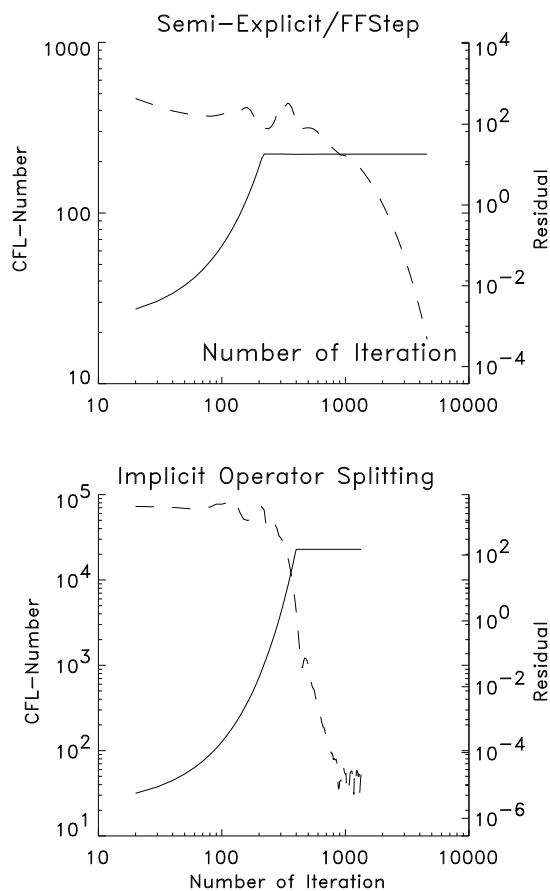}
}
\end{center}
{\vspace*{-0.0cm}} \caption [ ] {Free-fall of gas onto a black
hole surrounded by a disk. As in Fig. 7, the evolution of the
CFL-number and the residual versus number of iteration  are shown. The
semi-explicit method is apparently stable and converges to the
sought stationary solution, even when large time step sizes that
corresponds to CFL-number of 220 are used. On the other hand, 1000
iterations were sufficient to  obtain the same solution using the
implicit operator splitting approach (see Hujeirat \& Rannacher
2001).
  }
\end{figure}
\subsection{Weakly incompressible flows between two concentric spheres}
To
verify the capability of the method at modelling weakly incompressible flows,
 Taylor flows
between two concentric and rotating spheres has been tested (Fig. 14). This
is an ideal test problem, as the flow here is viscous (second
order diffusive operators are included), closed boundaries
(without external perturbations) and the corresponding equations
accept  strict time-independent solutions.\\
Fig. 13 shows the time-development of the CFL-number and the total
residual for 5-different solution procedures for searching steady
state configurations for Taylor flows  between two concentric
spheres. Using spherical geometry, the set of the 2D axi-symmetric
Navier-Stokes equations  are solved. The set consists of
 the three momentum equations, the continuity and the internal energy equations. The flow is assumed to
be adiabatic.\\
As boundary conditions, the inner sphere is set to rotate with
$\Omega = 5$, whereas the outer sphere has $\Omega = 0$.  The
density, temperature, radial and angular velocities are set to be
symmetric along the equator and along the polar axis, except the
angular velocity which assumes anti-symmetric conditions along the
axis of rotation. For the horizontal velocity, anti-symmetric
conditions are imposed both along the equator and along the axis
of rotation. On the inner and outer radius of the spheres, all
velocity-components are set to vanish. Initially, the flow between
\begin{figure*}
\begin{center}
{\hspace*{0.5cm}
\includegraphics*[width=6.5cm,bb=40 130 305 738 ,clip]{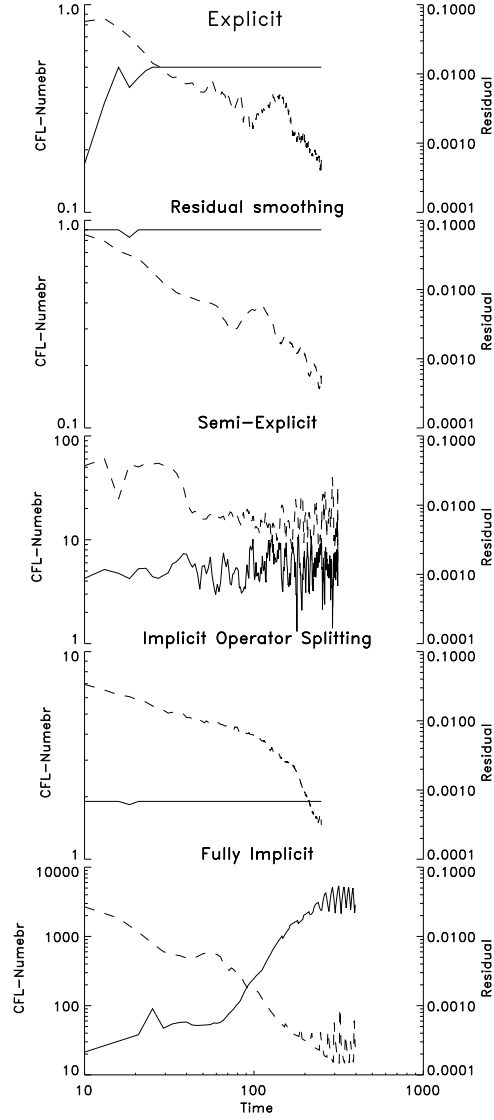}
}
\end{center}
{\vspace*{-0.0cm}} \caption [ ] { The development of the
CFL-number (left axis, solid line) and the
           total residual (right axis, dashed line) versus covered-time in normalized units of
          five different numerical methods  (from to top to bottom: normal explicit,
            residual smoothing,
           semi-explicit, implicit operator splitting (Hujeirat \& Rannacher 2001)
           and the fully implicit method).
           While the effective time covered in each run of these different methods
           is similar, the actual number of iteration is substantially different.
           The numerical problem here is to search stationary solutions for
           Taylor flow between two concentric spheres.
          The inner sphere has a radius $\,r_{in}\!=\!1\,$ and rotates with
    angular velocity $\,\Omega_{in}\!=\!5\,$, whereas the outer sphere is
         non-rotating and its radius is taken to be  $\,r_{out}\!=\!1.3\,$.
      $\theta=1$ and  $\,\eta\!=\!10^{-2}$ have been used in these
       calculations.
        The initial density and temperature are taken to be
       $\,\rho(r,\theta,t\!=\!0)\!=\!1\,$,
    and $\,T(r,\theta,t\!=\!0)\!=\!10^1\,$, respectively.
        The computational domain is
    $\,[1,1.3]\times[0,\pi/2]\,$ and consists of   $\,30\!\times\!50\,$  non-uniformly distributed
    tensor-product mesh.\\
  }
\end{figure*}
the two spheres has zero poloidal and toroidal velocities, so that
the rotational energy is injected into the flow through viscous
interaction with the inner boundary. In these test calculations,
 the viscosity coefficient $\eta = 10^{-2}$ and
the switch on/off parameter $\theta =1$ have been used.
\begin{figure}[htb]
\begin{center}
{\hspace*{-0.5cm}
\includegraphics*[width=6.75cm,bb=10 212 235 482,clip]{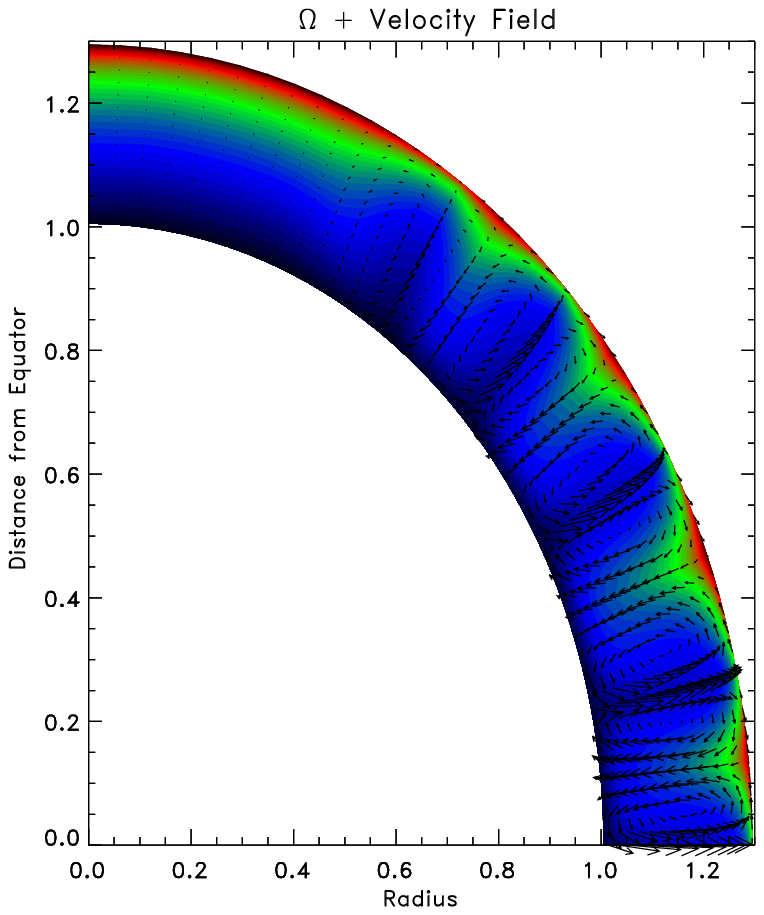}
}
\end{center}
{\vspace*{-0.0cm}} \caption [ ] {Steady state solutions of the
Taylor flow between two concentric spheres.
           Here the velocity field and the angular velocity
          (large-to-low values correspond to blue-to-red colors) are shown.
       The capability  of the methods to capture the formation of rotationally-driven
           multiple vortices near-equatorial
           region is obvious.
  }
\end{figure}
In the explicit case, the equations are solved according to Eq. 19.
For the semi-explicit procedure, we solve the HD-equations using
the block matrix formulation as described in Equation 20. The
implicit operator splitting approach is based in solving each of
the HD-equations implicitly. Here, the LGS method has been used in
the inversion procedure of each equation (Hujeirat \& Rannacher
2001). Unfortunately, while this method has been proven to be
robust for modeling compressible flows with open boundaries, it
fails to achieve large CFL-numbers in weakly incompressible flows
(Fig. 13). This indicates that pressure gradients in weakly
incompressible flows do not admit splitting, and therefore they
should be included in the solution procedure simultaneously at the new time level. 

Using the semi-explicit solution procedure, the CFL-numbers obtained in the present modelling
of Taylor flows  are, indeed, larger than unity (middle, Fig. 13), but they are not
impressively large  as we have  predicted theoretically. We may
attribute this inconsistency to three different effects:
 1) The flow considered here is weakly incompressible. This means that the acoustic perturbations have the
largest propagational speeds, which require that all pressure
effects should be included in the solution procedure
simultaneously on the new time level. Indeed, Fig. 15 shows that solving
the angular momentum only may relax these limitations and that
extremely large CFl-numbers can  be obtained.  
  2) The conditions imposed on the
boundaries are non-absorbing, and do not permit advection of
errors into regions exterior to the domain of calculation. 3) The
method requires probably additional improvements in order to
achieve large CFL-numbers. This could be done, for example, within
the context of the ``defect-correction'
iterative procedure, in which the block diagonal matrix $D_\mm{mod}$  is employed as a pre-conditioner.\\
\begin{figure}[t]
\begin{center}
{\hspace*{-0.5cm}
\includegraphics*[width=7.5cm,bb=47 185 295 315,clip]{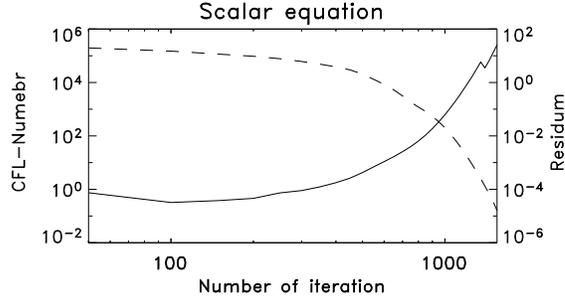}
}
\end{center}
{\vspace*{-0.0cm}} \caption [ ] {The semi-explicit method applied
to a scalar problem.
             As in Fig. 7, this  Figure shows the development of the CFL-number (left axis, solid line) and the
           residual (right axis, dashed line) of the angular momentum equation in two dimensions versus
            the number of iteration. As initial conditions we use
           the steady distributions of the physical variables that have been obtained
            from the simulations of the Taylor problem (see Fig. 14). This includes
            the velocity field, density, temperature and $\eta$. For the
            angular velocity we use $\Omega=0$ as initial condition.
  }
\end{figure}
In the final case (bottom, Fig. 13), the whole set of HD-equations is solved taking
into account all pressure terms in a fully implicit manner. Here we use the AFM for
solving the general block  matrix-equation which is locally described by Eq. 18. \\
For controlling the time step size in these calculations, we have
adopted the following description: $\delta t =
\alpha_0\eps/\max(RHS_{j,k})$, where $\eps = {\min}(\Delta X_j,
\Delta Y_k),$ and where $\alpha_0$ is a constant of order unity.
The maximum and minimum functions here are set to run over the
whole number of grid points. \\
Worth noting is the difference between the terminal values of the
residual in the semi-explicit and the fully implicit cases. After
a certain number of iteration, the residual in the former case
does not decrease, and instead it shows strong, but still limited
variations with increasing the number of iteration. Although a
CFL-number of order 10 is achieved in this test calculations, the
origin and magnitudes of these variations are not completely
clear. However, taking into account that these strong variations
do not show up in the fully implicit case, we may conclude that
their origin is connected to the incomplete inclusion of the
source terms in the coefficient matrix.
\section{Summary}
  
{
In this paper a strategy for modifying explicit methods and constructing a class
of semi-explicit methods is presented. We have shown that this modification enhance the
robustness and enlarge the range of application of explicit methods,
and in particular,  it enables their use for searching quasi-stationary solutions
for the Euler and Navier-Stokes equations in multi-dimensions.

\begin{table}[t]
\caption{The three different numerical approaches and their properties}
\begin{center}
\hspace*{-0.75cm}
\begin{tabular}{p{2.4cm}|p{3.75cm}p{4.0cm}p{3.0cm}}
  & Explicit & Semi-Explicit & Implicit \\ \hline
Stability: & stable within $~~~~~~~~~~$ $CFL<1$ 
           & absolutely stable 
           & absolutely stable \\
Convergence speed\footnotemark[1]:
           & limited by $~~~~~~~~$ $CFL<1$ 
           & faster than explicit, slower than implicit  
           & unlimited by the CFL condition \\
Efficiency\footnotemark[2]:  & highly efficient  
             & efficient 
             & inefficient   \\
Robustness:  & very limited         
             & robust, but can be made highly robust\footnotemark[3] 
             & robust \\
Programming efforts:  &  easy                                
             & easy                              
             & relatively difficult \\
\end{tabular}
\end{center}
\end{table}
\footnotetext[1]{This applies for diffusion dominated flows, and within the regime of stability of the method.
The discretization method used is assumed to be consistent/compatible with the continuous formulation.  
Semi-explicit methods converge relatively slowly, though not directly affected by the CFL-condition. }
\footnotetext[2]{A measure for the total computing operations per time step, provided the flow is time-dependent.
                 This measure may reverse if the total number of computing operation is of interest, and if 
                  quasi-stationary solutions are sought.}
\footnotetext[3]{Reducing the semi-explicit method into explicit when modelling strongly time-dependent flows, and
                 enhancing the degree of implicitness when modelling radiative, MHD and/or weakly/strongly incompressible flows
                 through incorporating additional non-diagonal entries (see Fig. 1).} 
The new strategy is based on reformulating explicit methods in
matrix form, i.e, transforming explicit methods from scalar into
matrix-vector problems. Thus, solving the set of time-dependent HD
or MHD equations would require solving  a matrix equation of the form:
$Aq=b$. Explicit methods in this case relies on approximating and
replacing the matrix $A$  by the most simple matrix in algebra:
the identity matrix I. This solution-procedure may converge, if
the time step size is sufficiently small, or equivalently, if the
entries of the matrix $I/\delta t$ are sufficiently large, so that
all off-diagonal elements of $A$ can be safely neglected. As we
have shown in Sec. 2, this yields a condition similar 
to the requirement that  $CFL <1$.

Unlike normal explicit methods, in which inclusion of diffusion
generally causes further limitations on the time step size, diffusion
operators in the semi-explicit formulation  pronounce the diagonal
dominance and enhances the stability of the inversion procedure,
irrespective of the dimensionality of the problem. However, 
the convergence rates of semi-explicit methods remain lower
than  that of their fully implicit counterparts.

A relevant question which arises here is: how efficient are
semi-explicit methods compared to explicit, and if they are more favorable
than explicit methods when modelling time-dependent flows? \\
We note that in applying the semi-explicit method to a scalar
equation, approximately $N$ additional algebraic operations are required for
calculating the diagonal elements, and roughly $N$ division operations.
Thus, applying the semi-explicit approach to system of equations of  $M$
variables, the additional algebraic operations  required scale as: 
$2\times N\times M$. If we take into account the addition global
iterations required per each time step to assure convergence,
 we conclude that the semi-explicit methods are, indeed,  less
efficient than their explicit counterparts, provided the flow
is strongly time-dependent. On the other hand, since the additional
algebraic operations increases linearly with the total number of
variables, the associated computational load   
is likely to be  small compared to  calculating the RHS of Eq. 7 or 16.
Furthermore, the algorithm can be optimized further, in such a manner that 
the relevant routines, in which the corresponding additional operations are performed,
are switched off if $CFL< 1 $. In this case, the semi-explicit approach degenerates 
into  an explicit method when the underlying flow is strongly time-dependent. 
This optimization procedure is particularly useful when modelling propagation of waves 
and moving shocks. The latter phenomena are non-linear features
of HD-flows, and the corresponding equations do not admit stationary solutions.
In this case, the time step size should be chosen sufficiently small in order to capture
shock profiles accurately,  irrespective of whether the method
used is explicit or implicit. For modelling such flows, explicit
methods are superior over implicit, provided that the flow is
isothermal, adiabatic or polytropic. In the case that other
physical and chemical processes are concerned, which generally
operate on much shorter time scales than the dynamical time scale,
modifying explicit methods into semi-explicit or strongly implicit is
unavoidable. \\
In Table 1, the main properties of the new semi-explicit method,
compared to explicit and implicit methods are listed.

Additionally, we have shown that the residual smoothing approach
accelerates the convergence of both explicit and semi-explicit
methods. \\ \\

Acknowledgement: This research has been partially supported by SFB 359 ``Reactive Str\"omungen,
 Diffusion und Transport" of the University of Heidelberg,  
Strasburg Observatory and the faculty of physics of Basel University. 
I thank the anonyous referee for her/his constructive and valuable suggestions, which
improved the clarity and readability of the paper significantly. 


\end{document}